\documentclass[sigplan,screen,authorversion]{acmart}
\setcopyright{acmlicensed}
\acmPrice{15.00}
\acmDOI{10.1145/3314221.3314645}
\acmYear{2019}
\copyrightyear{2019}
\acmISBN{978-1-4503-6712-7/19/06}
\acmConference[PLDI '19]{Proceedings of the 40th ACM SIGPLAN Conference on Programming Language Design and Implementation}{June 22--26, 2019}{Phoenix, AZ, USA}
\acmBooktitle{Proceedings of the 40th ACM SIGPLAN Conference on Programming Language Design and Implementation (PLDI '19), June 22--26, 2019, Phoenix, AZ, USA}

\usepackage{subcaption}
\usepackage[hypcap=true]{caption}
\usepackage{boxedminipage}
\usepackage{array} %
\usepackage{siunitx}
\usepackage{listings}
\usepackage{bold-extra}
\usepackage{graphicx}
\usepackage{tabularx}
\usepackage{multirow}
\usepackage{booktabs}
\usepackage{amsmath}
\usepackage{amsthm}
\usepackage{etoolbox}
\usepackage{enumitem} %
\usepackage[linesnumbered,ruled,vlined]{algorithm2e}
\SetAlFnt{\small\sf}

\makeatletter
\def\NAT@def@citea{\def\@citea{\NAT@separator{\,}}}
\makeatother

\lstdefinelanguage{JavaScript}{
  keywords={let,for,if,assert},
  keywordstyle=\bfseries,
  ndkeywords={},
  ndkeywordstyle=\bfseries,
  sensitive=false,
  comment=[l]{//},
  morecomment=[s]{/*}{*/},
  commentstyle=\ttfamily,
  stringstyle=\ttfamily,
  numberstyle=\scriptsize,
  morestring=[b]',
  morestring=[b]"
}

\theoremstyle{definition}
\newtheorem{defn}{Definition}

\newcommand{\strconcat}{\mathbin{\lstinline|++|}}

\DeclareMathOperator{\genin}{\boxdot}

\newcommand{\eqncode}[1]{\ensuremath{\text{\lstinline|#1|}}}
\newcommand{\lang}[1]{\mathcal{L}(#1)}
\newcommand{\clang}[1]{\mathcal{L}_c(#1)}
\newcommand{\capture}[1]{\mathcal{C}_{#1}}

\newcommand{\oldtext}[1]{}

\def\Snospace~{\S{}}

\renewcommand{\algorithmautorefname}{Algorithm}

\expandafter\patchcmd\csname \string\lstinline\endcsname{%
  \leavevmode
  \bgroup
}{%
  \leavevmode
  \ifmmode\hbox\fi
  \bgroup
}{}{%
  \typeout{Patching of \string\lstinline\space failed!}%
}

\begin{document}

\title[Sound Regular Expression Semantics for DSE of JavaScript]{Sound Regular Expression Semantics for Dynamic Symbolic Execution of JavaScript}

\author{Blake Loring}
\affiliation{
  \position{}
  \department{Information Security Group}
  \institution{\hspace*{-2mm}Royal Holloway, University of London\hspace*{-2mm}}
  \country{United Kingdom}
}
\email{blake.loring.2015@rhul.ac.uk}

\author{Duncan Mitchell}
\affiliation{
  \position{}
  \department{Department of Computer Science}
  \institution{\hspace*{-2mm}Royal Holloway, University of London\hspace*{-2mm}}
  \country{United Kingdom}
}
\email{duncan.mitchell.2015@rhul.ac.uk}

\author{Johannes Kinder}
\affiliation{
  \position{}
  \department{Research Institute CODE}
  \institution{Bundeswehr University Munich}
  \country{Germany}
}
\email{johannes.kinder@unibw.de}

\begin{abstract}
  Support for regular expressions in symbolic execution-based tools for test
  generation and bug finding is insufficient. Common aspects of mainstream
  regular expression engines, such as backreferences or greedy matching, are
  ignored or imprecisely approximated, leading to poor test coverage or missed
  bugs.
  In this paper, we present a model for the complete regular expression language
  of ECMAScript 2015 (ES6), which is sound for dynamic symbolic execution of the
  \lstinline|test| and \lstinline|exec| functions.  We model regular expression operations
  using string constraints and classical regular expressions and use a
  refinement scheme to address the problem of matching precedence and
  greediness.
  We implemented our model in ExpoSE, a dynamic symbolic execution engine for
  JavaScript, and evaluated it on over 1,000 Node.js packages containing regular
  expressions, demonstrating that the strategy is effective and can
  significantly increase the number of successful regular expression queries and
  therefore boost coverage.
\end{abstract}



\maketitle

\section{Introduction}

Regular expressions are popular with developers for matching and substituting
strings and are supported by many programming languages.
For instance, in JavaScript, one can write \lstinline|/goo+d/.test(s)| to test 
whether the string value of \lstinline|s| contains  
\lstinline!"go"!, 
followed by one or more occurrences of 
\lstinline!"o"! and a final 
\lstinline!"d"!.
Similarly, \lstinline|s.replace(/goo+d/,"better")| evaluates to a new string
where the first such occurrence in \lstinline|s| is replaced with the string
\lstinline!"better"!.

Several testing and verification tools include some degree of
support for regular expressions because they are so common~\cite{li2009reggae,
  veanes2010rex, Saxena:2010:SEF:1849417.1849985, TrinhCJ14, Loring:2017}.
In particular, SMT (satisfiability modulo theory) solvers now often support 
theories for strings and classical regular
expressions~\cite{Liang:2014:DTS:2735050.2735105, Liang:2015:DPR:2962076.2962089, norn, Abdulla:2017:FCF:3062341.3062384, TrinhCJ14,Zheng:2013:ZZS:2491411.2491456, ZhengGSTBDZ17, bjorner2012smt, DBLP:conf/cav/ZhengGSTDZ15, DBLP:journals/fac/FuPBL13},
which allow expressing constraints such as 
$\lstinline|s| \in \lang{\lstinline|goo+d|}$ for the \lstinline|test|
example above.
Although any general theory of strings is
undecidable~\cite{bjornerStringManipulation}, many string constraints are
efficiently solved by modern SMT solvers.

SMT solvers support regular expressions in the language-theoretical
sense, but ``regular expressions'' in programming languages like Perl or
JavaScript---often called \textit{regex}, a term we also adopt in the remainder of this paper---are not limited to representing regular
languages~\cite{Aho:1991:AFP:114872.114877}.
For instance, the expression \lstinline|/<(\w+)>.*?<\/\1>/| parses any pair of matching XML tags, which is a context-sensitive language (because the tag is an arbitrary string that must appear twice).
Problematic features that prevent a translation of regexes to the word problem
in regular languages include capture groups (the parentheses around
\lstinline|\w+| in the example above), backreferences (the \lstinline|\1|
referring to the capture group), and greedy/non-greedy matching precedence of
subexpressions (the \lstinline|.*?| is non-greedy).
In addition, any such expression could also be included in a lookahead
\lstinline|(?=)|, which effectively encodes intersection of context sensitive
languages.
In tools reasoning about string-manipulating
programs, these features are usually ignored or imprecisely approximated. This
is a problem, because they are widely used,
as we demonstrate in \autoref{sec:survey}.

In the context of dynamic symbolic execution (DSE) for test generation, this lack of
support can lead to loss of coverage or missed bugs where constraints would have
to include membership in non-regular languages. The difficulty arises from the
typical mixing of constraints in path conditions---simply \textit{generating} a
matching word for a standalone regex is easy (without lookaheads).
To date, there has been only limited progress on this problem, mostly addressing
immediate needs of implementations with approximate solutions, e.g., for capture
groups~\cite{Saxena:2010:SEF:1849417.1849985} and
backreferences~\cite{Scott2015,Loring:2017}. However, neither matching
precedence nor lookaheads have been addressed before. 

In this paper, we propose a novel scheme for supporting ECMAScript
regex in dynamic symbolic execution and show that it is effective
in practice.
We rely on the specification of regexes and their associated 
methods in ECMAScript 2015 (ES6). However, our methods and findings should be
easily transferable to most other existing implementations.
In particular, we make the following contributions:
\begin{itemize}
\item We fully model ES6 regex in terms of classical regular
  languages and string constraints~(\autoref{sec:encoding}) and
  cover several aspects missing from previous
  work \cite{Saxena:2010:SEF:1849417.1849985, Scott2015, Loring:2017}. We
  introduce the notion of a \textit{capturing language} to make the problem of
  matching and capture group assignment self-contained.
\item We introduce a counterexample-guided abstraction refinement (CEGAR) scheme
  to address the effect of greediness on capture
  groups~(\autoref{sec:refinement}), which allows us to deploy our model in DSE
  without sacrificing soundness for under-approximation.
\item We present the first systematic study of JavaScript regexes,
  examining feature usage across 415,487 packages from
  the NPM software repository. We show that non-regular features are widely 
  used~(\autoref{sec:survey}).
\end{itemize}

In the remainder of the paper we review ES6 regexes~(\autoref{sec:background}) and present an overview of our approach
by example (\autoref{sec:approach}). We then detail our regex model using a novel formulation~(\autoref{sec:encoding}), and we propose a
CEGAR scheme to address matching precedence (\autoref{sec:refinement}).
We discuss an implementation of the model as part of the ExpoSE symbolic
execution engine for JavaScript (\autoref{s:impl}) and evaluate its practical
impact on DSE (\autoref{sec:eval}).
Finally, we review related work~(\autoref{sec:relatedwork}) and
conclude~(\autoref{sec:conclusion}).

\section{ECMAScript Regex}
\label{sec:background}

We review the ES6 regex specification, focusing on differences to classical regular
expressions. We begin with the regex API and its matching
behavior (\autoref{sec:api}) and then explain capture groups (\autoref{s:bgcg}),
backreferences (\autoref{s:br}), and operator precedence (\autoref{s:op-prec}).
ES6 regexes are comparable to those of other languages but lack
Perl's recursion and lookbehind and do not require POSIX-like longest matches.

\subsection{Methods, Anchors, Flags}
\label{sec:api}

ES6 regexes are \lstinline|RegExp| objects, created  
from literals or the \lstinline!RegExp! constructor.
\lstinline!RegExp! objects have two methods, \lstinline|test| and
\lstinline|exec|, which expect a string argument; String objects offer the
\lstinline|match|, \lstinline|split|, \lstinline|search| and 
\lstinline|replace| methods that expect a \lstinline|RegExp| argument.

A regex accepts a string if any portion of the string
matches the expression, i.e., it is implicitly surrounded by wildcards.
The relative position in the string can be controlled with
\textit{anchors}, with \textasciicircum{} and \$ matching the start and end,
respectively.

\textit{Flags} in regexes can modify the behavior of matching
operations.
The \textit{ignore case} flag \lstinline|i| ignores character cases when
matching.
The \textit{multiline} flag \lstinline|m| redefines anchor characters
to match either the start and end of input or newline characters.
The \textit{unicode} flag \lstinline|u| changes how unicode literals are escaped
within an expression. 
The \textit{sticky} flag \lstinline|y| forces matching to start at
\mbox{\lstinline|RegExp.lastIndex|,} which is updated with the index of the
previous match. Therefore, \lstinline|RegExp| objects become stateful as seen in
the following example:
\begin{lstlisting}
  r = /goo+d/y;
  r.test("goood"); // true;  r.lastIndex = 6
  r.test("goood"); // false; r.lastIndex = 0
\end{lstlisting}
The meaning of the \textit{global} flag \lstinline|g| varies. It extends the
effects of \lstinline|match| and \lstinline|replace| to include all matches on
the string and it is equivalent to the sticky flag for the \lstinline|test| and
\lstinline|exec| methods of \mbox{\lstinline|RegExp|.}

\subsection{Capture Groups}\label{s:bgcg}

Parentheses in regexes not only change operator precedence (e.g.,
\lstinline|(ab)*| matches any number of repetitions of the string 
\lstinline!"ab"! while \lstinline|ab*| matches the character \lstinline!"a"! 
followed by any number of repetitions of the character \lstinline!"b"!)
but also create \textit{capture groups}.
Capture groups are implicitly numbered from left to right by order of
the opening parenthesis. For example, \lstinline!/a|((b)*c)*d/! is numbered as
$\lstinline!/a|(!^1\lstinline!(!^2\lstinline!b)*c)*d/!$.
Where only bracketing is required, a non-capturing group can be created by 
using the syntax $\lstinline!(?:! \ldots \lstinline!)!$.

For regexes, capture groups are important because the
regex engine will record the \textit{most recent} substring 
matched
against each capture group. Capture groups can be referred to from within the
expression using backreferences (see \autoref{s:br}). The last matched
substring for each capture group is also returned by some of the API methods.
In JavaScript, the return values of \lstinline|match| and \lstinline|exec| are arrays, with the 
whole match at index $0$ (the implicit capture group $0$), and the last matched instance of the $i^{\text{th}}$ 
capture group at index $i$.
In the example above, 
\lstinline+"bbbbcbcd".match(/a|((b)*c)*d/)+
will evaluate to the array \lstinline|["bbbbcbcd", "bc", "b"]|.

\subsection{Backreferences}\label{s:br}

A \textit{backreference} in a regex refers to a numbered capture
group and will match whatever the engine last matched the capture group
against. 
In general, the addition of backreferences to regexes makes the
accepted languages non-regular~\cite{Aho:1991:AFP:114872.114877}.

Inside quantifiers (Kleene star, Kleene plus, and other repetition operators),
the string matched by the backreference can change across multiple matches.
For example, the regex \lstinline!/((a|b)\2)+/! can match the 
string \lstinline!"aabb"!, with the backreference \lstinline!\2! being matched 
twice: the first time, the capture group contains \lstinline!"a"!, the second 
time it contains \lstinline!"b"!.
This logic applies recursively, and it is possible for backreferences to in 
turn be part of outer capture groups.

\begin{table}[t]
\caption{Regular expression operators, separated by classes of precedence.}
\label{tbl:operators}
\small
\begin{tabularx}{\columnwidth}{p{13mm}Xp{15mm}}
  \toprule
  \bf Operator & \bf Name & \bf Rewriting \\ 

  \midrule

  $\lstinline!(!r\lstinline!)!$ & Capturing parentheses  
  \\
  $\backslash n$ & Backreference 
  &
  \\
  $\lstinline!(?:!r\lstinline!)!$ & Non-capturing parentheses 
  \\

  $\lstinline|(?=|r\lstinline|)|$ & Positive lookahead
  \\

  $\lstinline|(?!|r\lstinline|)|$ & Negative lookahead
  \\

  $\lstinline!$\backslash$ b!$ & Word boundary
  \\
  
  $\lstinline!$\backslash$ B!$ & Non-word boundary
  \\

  \midrule
  $r\,\lstinline!*!$ & Kleene star
  \\
  $r\,\lstinline!*?!$ & Lazy Kleene star
  \\
  $r\,\lstinline!+!$ & Kleene plus 
  & $r\,\lstinline!*!r$
  \\
  $r\,\lstinline!+?!$ & Lazy Kleene plus 
  & $r\,\lstinline!*?!r$
  \\
  $r\,\lstinline!\{$m$,$n$\}!$ & Repetition
  & $r^n \lstinline!|! \ldots \lstinline!|! r^m$
  \\
  $r\,\lstinline!\{$m$,$n$\}?!$ & Lazy repetition
  & $r^m \lstinline!|! \ldots \lstinline!|! r^n$
  \\
  \midrule
  $r\lstinline!?!$ & Optional
  & $r \lstinline!|! \epsilon$
  \\
  $r\lstinline!??!$ & Lazy optional
  & $\epsilon \lstinline!|! r$
  \\
  \midrule
  $r_1r_2$ & Concatenation
  \\
  \midrule
  $r_1 \lstinline!|! r_2$ & Alternation 
  \\
  \bottomrule
\end{tabularx}
\end{table}

\subsection{Operator Evaluation}
\label{s:op-prec}

We explain the operators of interest for this paper in \autoref{tbl:operators};
the implementation described in \autoref{s:impl} supports the full ES6
syntax~\cite{ecma-spec}. Some operators can be rewritten into semantically
equivalent expressions to reduce the number of cases to handle (shown in the
\textbf{Rewriting} column).

Regexes distinguish between \textit{greedy} and
\textit{lazy} evaluation. Greedy operators consume as many characters
as possible such that the entire regular expression still matches; lazy
operators consume as few characters as possible.
This distinction---called \textit{matching precedence}---is 
unnecessary for classical regular languages, but does affect the assignment 
of capture groups and therefore backreferences.

\textit{Zero-length assertions} or \textit{lookarounds} do not consume any characters but still restrict the accepted word,
enforcing a language intersection.
Positive or negative \textit{lookaheads} can contain any regex,
including capture groups and backreferences.
In ES6, \textit{lookbehind} is only available through \lstinline|\b| (word
boundary), and \lstinline|\B| (non-word boundary), which are commonly used to
only (or never) match whole words in a string.

\section{Overview}
\label{sec:approach}
In an overview of our approach, we now define the word problem for
regex~(\autoref{s:word_problem}) and how it arises in
DSE~(\autoref{s:problem-dse}). We introduce our model for regex by example 
(\autoref{sec:overview_model}) and explain how to eliminate
spurious solutions by refinement (\autoref{sec:overview_refinement}).

\subsection{The Word Problem and Capturing Languages}
\label{s:word_problem}
For any given classical regular expression $r$, we write $w \in \lang{r}$
whenever $w$ is a word within the (regular) language generated by $r$.
For a regex $R$, we also need to record values of capture
groups within the regex.
To this end, we introduce the following notion:
\begin{defn}[Capturing Language] \label{def:CL} The \textit{capturing language}
  of a regex $R$, denoted $\clang{R}$, is the set of tuples
  $(w, \capture{0}, \ldots, \capture{n})$ such that $w$ is a word of the
  language of $R$ and each $\capture{0}, \ldots, \capture{n}$ is the
  substring of $w$ matched by the corresponding numbered capture group in $R$.
\end{defn}%
\noindent A word $w$ is therefore matched by a regex $R$ if and only if
$\exists \capture{0}, \ldots, \capture{n}: (w, \capture{0}, \ldots, \capture{n})
\in \clang{R}$.  It is not matched if and only if
$\forall \capture{0}, \ldots, \capture{n}: (w, \capture{0}, \ldots, \capture{n})
\not\in \clang{R}$.  For readability, we will usually omit quantifiers for
capture variables where they are clear from the context.

\subsection{Regex In Dynamic Symbolic Execution}\label{s:problem-dse}

The code in \autoref{fig:re_js_example} parses numeric arguments between XML
tags from its input variable \lstinline|args|, an array of strings. The regex
in line 4 breaks each argument into two capture groups, the tag and
the numeric value (\lstinline|parts[0]| is the entire match). When the tag is
``\lstinline|timeout|'', it sets the \lstinline|timeout| value accordingly
(lines 6--7).
On line 12, a runtime assertion checks that the timeout value is truly
numeric after the arguments have been processed. The assertion can
fail because the program contains a bug: the regex in line 4 uses a
Kleene star and therefore also admits the empty string as the number to set, 
and JavaScript's dynamic type system will allow setting \lstinline|timeout| to 
the empty string.
\begin{lstlisting}[float,caption=Code example using regex., ,
mathescape=false,basicstyle=\ttfamily\small,label=fig:re_js_example,numbers=left,xleftmargin=2em]
let timeout = '500'; 
for (let i = 0; i < args.length; i++) {
  let arg = args[i];
  let parts = /<(\w+)>([0-9]*)<\/\1>/.exec(arg);
  if (parts) {
    if (parts[1] === "timeout") {
      timeout = parts[2];
    }
    ...
  }
}
assert(/^[0-9]+$/.test(timeout) == true);
\end{lstlisting}

DSE finds such bugs by systematically enumerating paths, including the failure
branches of assertions~\cite{dart}. 
Starting from a concrete run with input, say, \lstinline|args[0] = "foo"|, the
DSE engine will attempt to build a \textit{path condition} that encodes the
branching decisions in terms of the input values. It then attempts to
systematically flip clauses in the path condition and query an SMT solver to
obtain input assignments covering different paths. This process repeats forever
or until all paths are covered (this program has an unbounded number of paths as
it is looping over an input string).

Without support for regex, the DSE engine will \textit{concretize}
\lstinline|arg| on the call to \lstinline|exec|, assigning the concrete result
to \lstinline|parts|. With all subsequent decisions therefore concrete, the path
condition becomes $\mathit{pc} = \lstinline|true|$ and the engine will be
unable to cover more paths and find the bug.

Implementing regex support ensures that \lstinline|parts| is
\textit{symbolic}, i.e., its elements are represented as formulas during
symbolic execution. The path condition for the initial path thus becomes
$\mathit{pc} = (\lstinline|args[0]|, \capture{0}, \capture{1}, \capture{2}) \not\in \clang{R}$ where
$R = \lstinline[mathescape=false]|<(\w+)>([0-9]*)<\/\1>|$. Negating the only
clause and solving yields, e.g., $\lstinline|args[0]| =
\lstinline|"<a>0</a>"|$. DSE then uses this input assignment to cover a second
path with
$\mathit{pc} = (\lstinline|args[0]|, \capture{0}, \capture{1}, \capture{2}) \in \clang{R} \wedge \capture{1}
\neq \lstinline|"timeout"|$. Ne\-gating the last clause yields, e.g.,
``\lstinline|<timeout>0</timeout>|'', entering line 7 and making
\lstinline|timeout| and therefore the assertion symbolic. This leads to
$pc = (\lstinline|args[0]|, \capture{0}, \capture{1}, \capture{2}) \in \clang{R} \wedge \capture{1} =
\lstinline|"timeout"| \wedge (\capture{2}, {\mathcal{C}'_0}) \in \clang{\lstinline|^[0-9]+$\$$|}$, which, after
negating the last clause, triggers the bug with the input
``\lstinline|<timeout></timeout>|''.

\subsection{Modeling Capturing Language Membership}
\label{sec:overview_model}
Capturing language membership constraints in the path condition cannot be
directly expressed in SMT. We model these in terms of classical
regular language membership and string constraints.
For a given ES6 regex $R$, we first rewrite 
$R$ (see \autoref{tbl:operators}) in atomic terms only, i.e., 
\lstinline{|}, \lstinline{*}, capture groups, backreferences, lookaheads, and anchors.
For consistency with the JavaScript API, we also introduce the outer
capture group $\capture{0}$.
Consider the regex 
$R = \lstinline!(?:a|(b))$\backslash$1$$!$.
After preprocessing, the capturing language membership problem becomes
\[(w, \capture{0}, \capture{1}) \in
\clang{\lstinline[mathescape=false]!(?:.|\\n)*?((?:a|(b))\\1)(?:.|\\n)*?!},\]
a generic rewriting that allows for characters to precede and follow the
match in the absence of anchors.

We recursively reduce capturing language membership to regular membership.
To begin, we translate the purely regular Kleene stars and the outer
capture group to obtain
\begin{align*}
(w, \capture{0}, \capture{1}) \in & \clang{R} \implies w = w_1 \strconcat w_2 \strconcat w_3  \\
&\land w_1 \in \mathcal{L}(\lstinline[mathescape=false]!(:?.|\\n)*?!) \\
&\land (w_2, \capture{1}) \in \clang{\lstinline[mathescape=false]!(?:a|(b))\\1!} \land \; \capture{0} = w_2 \\
  &\land w_3 \in \mathcal{L}(\lstinline[mathescape=false]!(:?.|\\n)*?!),
\end{align*}
where $\;\strconcat$ is string concatenation.
We continue by decomposing the regex until there are only purely regular terms or standard string constraints.
Next, we translate the nested capturing language constraint
\begin{multline*}
  (w_2, C_1) \in \clang{\lstinline!(?:a|(b))\\1!} \implies \\
  \quad \; w_2 = w'_1 \strconcat w'_2 \land (w'_1, \capture{1}) \in \clang{\lstinline!a|(b)!} \land (w'_2) \in \clang{\lstinline[mathescape]!$\backslash$1!}.
\end{multline*}
When treating the alternation, either the left is satisfied and the capture 
group becomes undefined (which we denote as $\varnothing$), or the right is 
satisfied and the capture is locked to the match, which we model as
\[(w'_1 \in \lang{a} \land \capture{1}=\varnothing) \lor (w'_1 \in \lang{b} \land \capture{1} = w'_1).\]
Finally we model the backreference, which is case dependent on whether the capture group it refers to is defined or not:
\[(\capture{1} = \varnothing \implies w'_2 = \epsilon) \land (\capture{1} \neq \varnothing \implies w'_2 = \capture{1}).\]
Putting this together, we obtain a model for $R$:
\begin{align*}
&(w, \capture{0}, \capture{1}) \in \clang{R} \implies w = w_1 \strconcat w'_1 \strconcat w'_2 \strconcat w_3 \\
&\land \capture{0} = w'_1 \strconcat w'_2 \\
&\land \big( (w'_1 \in \lang{a} \land \capture{1} = \varnothing) \lor (w'_1 \in \lang{b} \land \capture{1} = w'_1) \big) \\
&\land (\capture{1} = \varnothing \implies w'_2 = \epsilon) \land (\capture{1} \neq \varnothing \implies w'_2 = \capture{1}) \\
&\land w_1 \in \mathcal{L}(\lstinline!(:?.|\\n)*?!) \land w_3 \in \mathcal{L}(\lstinline!(:?.|\\n)*?!).
\end{align*}

\subsection{Refinement}
\label{sec:overview_refinement}
Because of matching precedence (greediness), these models permit
assignments to capture groups that are impossible in real executions.
For example, we model $\lstinline[mathescape=false]!/^a*(a)?$/!$ as
\begin{multline*}
(w, \capture{0}, \capture{1}) \in \clang{\lstinline[mathescape=false]!/^a*(a)?$/!} \implies 
  w = w_1 \strconcat w_2 \\
  \land w_1 \in \lang{\lstinline!a*!} 
    \land w_2 \in \lang{\lstinline!a! | \epsilon} 
  \land \capture{0} = w \land \capture{1} = w_2.
\end{multline*}
This allows $\capture{1}$ to be either $\lstinline!a!$ or the empty string
$\epsilon$, i.e., the tuple 
$(\lstinline!"aa"!, \lstinline!"aa"!, \lstinline!"a"!)$ would be a spurious member of the capturing language under our model.
Because \lstinline!a*! is \textit{greedy}, it will always consume both
characters in the string \lstinline!"aa"!; therefore, \lstinline!(a)?! can only
match $\epsilon$.
This problem posed by \textit{greedy} and \textit{lazy} operator semantics 
remains unaddressed by previous 
work~\cite{Saxena:2010:SEF:1849417.1849985, TrinhCJ14, Loring:2017, Scott2015}.
To address this, we use a counterexample-guided abstraction refinement scheme
that validates candidate assignments with an ES6-compliant matcher.
Continuing the example, 
the candidate element $(\lstinline!"aa"!, \lstinline!"aa"!, \lstinline!"a"!)$ is
validated by running a concrete matcher on the string $\lstinline!"aa"!$, which
contradicts the candidate captures with $\capture{0}=\lstinline!"aa"!$ and
$\capture{1}=\epsilon$.
The model is refined with the counter-example to the following:
\begin{align*}
  w = & w_1 \strconcat w_2 \\
    &\land w_1 \in \lang{\lstinline!a*!} \land w_2 \in \lang{\lstinline!a! | \epsilon} 
    \land \capture{0} = w \land \capture{1} = w_2 \\
    &\land \big( w = \lstinline!"aa"! \implies \left(\capture{0} = 
      \lstinline!"aa"! \land \capture{1} = \epsilon \right)\big).
\end{align*}%
We then generate and validate a new candidate $(w, \capture{0}, \capture{1})$
and repeat the refinement until a satisfying assignment passes the concrete
matcher.

\section{Modeling ES6 Regex}
\label{sec:encoding}

We now detail the process of modeling capturing languages. After preprocessing a
given ES6 regex $R$ to $R'$ (\autoref{s:prelim}), we model
constraints $(w, \capture{0}, \ldots, \capture{n}) \in \clang{R'}$ by
recursively translating terms in the abstract syntax tree (AST) of $R'$ to
classical regular language membership and string constraints
(\autoref{s:modeling-ops}--\ref{s:brefs}).
Finally, we show how to model negated constraints
$(w, \capture{0}, \ldots, \capture{n}) \not\in \clang{R'}$
(\autoref{s:complement}).

\subsection{Preprocessing}
\label{s:prelim}
 
For illustrative purposes, we make the concatenation $R_1R_2$ of terms
$R_1, R_2$ explicit as the binary operator $R_1 \cdot R_2$.
Any regex can then be split into combinations of atomic elements,
capture groups and backreferences (referred to collectively as \textit{terms},
in line with the ES6 specification~\cite{ecma-spec}), joined by explicit
operators.
Using the rules in \autoref{tbl:operators}, we rewrite any %
$R$ to an equivalent regex $R'$ containing only alternation,
concatenation, Kleene star, capture groups, non-capturing parentheses,
lookarounds, and backreferences.
We rewrite any remaining lazy quantifiers to their greedy equivalents, as the
models are agnostic to matching precedence (this is dealt with in
refinement).

Note that the rules for Kleene plus and repetition duplicate capture groups, %
e.g., rewriting $\lstinline!/(a){1,2}/!$ to
$\lstinline!/(a)(a)|(a)/!$ adds two capture
groups.
We therefore explicitly relate capture groups between the original and rewritten
expressions.
When rewriting a Kleene
plus expression $S+$ containing $K$ capture groups, $S\lstinline!*!S$ has
$2K$ capture groups. For a constraint of the form
$(\capture{1}, \ldots, \capture{K}) \in \clang{S+}$, the rewriting yields 
\[(\capture{0}, \capture{{1,1}}, \ldots, \capture{{K,1}}, \capture{{1,2}}, \ldots,
\capture{{K,2}}) \in \clang{S\lstinline!*!S}.\]
Since $S\lstinline!*!S$ contains two copies of $S$, $\capture{{i,j}}$ 
corresponds to the $i^\text{th}$ capture in the $j^\text{th}$ copy of $S$ in 
$S\lstinline!*!S$. We express the direct correspondence between captures as
\begin{align*}
  (w, \capture{0}, \capture{1}, &\ldots, \capture{K}) \in \clang{S+} \iff \\
  &(w, \capture{0}, \capture{{1,1}}, \ldots, \capture{{K,1}}, \capture{{1,2}}, 
    \ldots, \capture{{K,2}}) \in \clang{S\lstinline!*!S} \\
    \land &\forall i \in \{1,\ldots, K\}, \: \capture{i} = \capture{{i,2}}.
\end{align*}

For repetition, if $S\{m,n\}$ has $K$ 
capture groups, then 
$S' = S^n \mid \ldots \mid S^m$ has $\frac{K}{2}(n+m)(n-m+1)$ captures.
In $S'$, suppose we index our captures as $\capture{{i,j,k}}$ where 
$i \in \{1, \ldots, K\}$ is the index of the capture group in $S$, 
$j \in \{0, \ldots, n-m \}$ denotes which alternate the capture group is in (0 being the rightmost),
and $k \in \{0, \ldots, m+j-1 \}$ indexes the copies of $S$ within each
alternate.
Intuitively, we pick a single $x \in \{0,\ldots, n-m\}$ 
that corresponds to the first satisfied alternate. Comparing the 
assignment of captures in $S\{m,n\}$ to $S'$, we know that the value of the 
capture is the last possible match, so $\capture{i} = \capture{i, x, m+x-1}$ for 
all $i \in \{1, \ldots, K\}$.
Formally, this direct correspondence can be expressed as
\begin{align*}
  (w, &\capture{0}, \capture{1}, \ldots, \capture{K}) \in \clang{S\{m,n\}} 
    \iff \\
  &(w, \capture{0}, \capture{{1,0,0}}, \ldots, \capture{{K, n-m, n}}) \in 
    \clang{S^n \mid \ldots \mid S^m} \\
    \land &\exists x \in\{0, \ldots, n-m\} : \\
    &\big(
      (w, \capture{0}, \capture{{1,x,0}}, \ldots, \capture{{K, x, m+x-1}}) \in 
        \clang{S^{m+x}}\\
      &\:\:\land \forall x' > x, \: (w, \capture{0}, \capture{{1,x',0}}, 
        \ldots, \capture{{K, x', m+x'-1}}) \not\in \clang{S^{m+x'}} \\
      &\:\:\land \forall i \in \{1,\ldots, K\}, \: 
        \capture{i} = \capture{{i,x,m+x-1}} \big).\\[-4pt] %
\end{align*}

\subsection{Operators and Capture Groups}\label{s:modeling-ops}
\begin{table*}[tp]
\caption{Models for regex operators.}
\label{tbl:models}
\small
\begin{tabularx}{\textwidth}{m{28mm}p{14mm}X}
  \toprule
  \bf Operation & \bf $t$  & \bf Overapproximate Model for $(w, \capture{i},...,\capture{i+k}) \in \clang{t}$ \\

  \midrule

  Alternation &
  $t_1 \lstinline!|! t_2$ &
  {$\!\begin{aligned}
    &\big( (w, \capture{i},...,\capture{i+j}) \in \clang{t_1} 
    \land \capture{i+j+1} = ... = \capture{i+k} = \varnothing \big) \\
    \lor &\big( (w, \capture{i+j+1},...,\capture{i+k}) \in \clang{t_2} \land
    \capture{i} = ... = \capture{i+j} = \varnothing \big)\end{aligned}$} \\

  \midrule

  Concatenation &
  $t_1 \cdot t_2$ & 
  $w = w_1 \strconcat w_2 \land (w_1, \capture{i},...,\capture{i+j}) \in \clang{t_1} \land (w_2, \capture{i+j+1},...,\capture{i+k}) \in \clang{t_2} $ \\
  \midrule

  Backreference-free Quantification &
  $t_1\lstinline!*!$ & %
  {$\!\begin{aligned}
    w = w_1 \strconcat w_2 
    &\land w_1 \in \lang{\hat{t_1}\lstinline!*!}
    \land (w_2, \capture{i},...,\capture{i+k}) \in \clang{t_1\lstinline!|! \epsilon} \\
    &\land \big(w_2 = \epsilon \implies (w_1 = \epsilon \land \capture{i}=\ldots=\capture{i+k}=\varnothing)\big)
    \end{aligned}$} \\
  
  \midrule

  Positive Lookahead &
  $\lstinline!(?=$t_1$)! t_2$ &
  $(w, \capture{i},...,\capture{i+j}) \in \clang{\lstinline!$t_1$.*!}
    \land (w, \capture{i+j+1},...,\capture{i+k}) \in \clang{t_2}$ \\

  \midrule

  Negative Lookahead &
  $\lstinline|(!=$t_1$)| t_2$ &
  $(w, \capture{i},...,\capture{i+j}) \not\in \clang{\lstinline!$t_1$.*!}
    \land (w, \capture{i+j+1},...,\capture{i+k}) \in \clang{t_2}$ \\

  \midrule

  Input Start &
  $t_1$\lstinline|^| &
  $(w, \capture{i},...,\capture{i+k}) \in \clang{\lstinline!$t_1$!} \land (w, \capture{i},...,\capture{i+k}) \in \mathcal{L}(.*\langle)$\\

  \midrule

  Input Start (Multiline) &
  $t_1$\lstinline|^| &
  $(w, \capture{i},...,\capture{i+k}) \in \clang{\lstinline!$t_1$!} \land (w, \capture{i},...,\capture{i+k}) \in \mathcal{L}(.*\langle|\backslash n)$\\

  \midrule
  Input End &
  \lstinline[mathescape=false]|$|$t_1$ &
  $(w, \capture{i},...,\capture{i+k}) \in \clang{\lstinline!$t_1$!} \land (w, \capture{i},...,\capture{i+k}) \in \mathcal{L}(\rangle.*)$ \\

  \midrule
  Input End (Multiline) &
  \lstinline[mathescape=false]|$|$t_1$ &
  $(w, \capture{i},...,\capture{i+k}) \in \clang{\lstinline!$t_1$!} \land (w, \capture{i},...,\capture{i+k}) \in \mathcal{L}(\rangle|\backslash n.*)$ \\

  \midrule

  Word Boundary &
  $t_1 \backslash \lstinline!b! \; t_2$ &
  {$\!\begin{aligned}
   w &= w_1 \strconcat w_2
    \land (w_1, \capture{i},...,\capture{i+j}) \in \clang{t_1} 
    \land (w_2, \capture{i+j+1},...,\capture{i+k}) \in \clang{t_2} \\
    & \land \Big( \big( (w_1 \in \mathcal{L}(\texttt{.*}\backslash\texttt{W}) 
    \lor w_1 = \epsilon ) \land w_2 \in \mathcal{L}(\backslash \texttt{w.*}) \big) 
     \lor \big(w_1 \in \mathcal{L}(\texttt{.*}\backslash\texttt{w}) 
      \land (w_2 \in \mathcal{L}(\backslash\texttt{W.*}) \lor w_2=\epsilon) \big) \Big)
    \end{aligned}$} \\

  \midrule

  Non-Word Boundary &
  $t_1 \backslash \lstinline!B! \; t_2$ &
  {$\!\begin{aligned}
    w &=  w_1 \strconcat w_2
    \land (w_1, \capture{i},...,\capture{i+j}) \in \clang{t_1} 
    \land (w_2, \capture{i+j+1},...,\capture{i+k}) \in \clang{t_2} \\
    & \land \big( (w_1 \not\in \mathcal{L}(\texttt{.*}\backslash\texttt{W}) 
     \land w_1 \neq \epsilon) 
     \lor w_2 \not\in \mathcal{L}(\backslash \texttt{w.*}) \big) 
    \land \big(w_1 \not\in \mathcal{L}(\texttt{.*}\backslash\texttt{w}) 
      \lor (w_2 \not\in \mathcal{L}(\backslash\texttt{W.*}) \land w_2\neq\epsilon)\big)
    \end{aligned}$} \\

  \midrule

  Capture Group &
  $(t_1)$ &
  {$\!\begin{aligned}
    (w, \capture{i+1},..., \capture{i+k}) \in \clang{t_1}
    \land \capture{i} = w
  \end{aligned}$} \\

  \midrule

  Non-Capturing Group &
  $(\lstinline!?:!t_1)$ &
  {$\!\begin{aligned}
    (w, \capture{i},..., \capture{i+k}) \in \clang{t_1}
  \end{aligned}$} \\

  \midrule

  Base Case &
  $t$ regular &
  {$\!\begin{aligned}
      w \in \lang{t}
  \end{aligned}$} \\

  \bottomrule
\end{tabularx}
\end{table*}

Let $t$ be the next term to process in the AST of $R'$. If $t$ is capture-free and
purely regular, there is nothing to do in this step.
If $t$ is non-regular, it contains $k+1$ capture groups (with $k\!\geq\!-1$) numbered $i$ through $i+k$.
At each recursive step, we express membership of the capturing language
$(w, \capture{i}, ..., \capture{i+k}) \in \clang{t}$ through a model consisting
of string and regular language membership constraints, and a set of remaining
capturing language membership constraints for subterms of $t$.
Note that we record the locations of capture groups within the regex in the 
preprocessing step. When splitting $t$ into subterms $t_1$ and
$t_2$, capture groups $\capture{i}, \ldots, \capture{i+j}$ are contained in
$t_1$ and $\capture{i+j+1}, \ldots, \capture{i+k}$ are contained in $t_2$ for
some $j$.
The models for individual operations are given in \autoref{tbl:models}; we
discuss specifics of the rules below.

When matching an alternation \lstinline+|+, capture groups on the non-matching side
will be undefined, denoted by $\varnothing$, which is distinct from the empty
string $\epsilon$.

When modeling quantification $t = t_1*$, we assume $t_1$ does not contain
backreferences (we address this case in \autoref{s:brefs}).
In this instance, we model $t$ via the expression 
$\hat{t_1}\lstinline!*!t_1\lstinline!|!\epsilon$, where $\hat{t_1}$ is a regular 
expression corresponding to $t_1$, except each set of capturing parentheses is 
rewritten as a set of non-capturing parentheses.
In this way, $\hat{t_1}$ is regular (it is backreference-free by assumption).
However, $\hat{t_1}\lstinline!*!t_1\lstinline!|!\epsilon$ is not semantically 
equivalent to $t$: if possible, capturing groups must be satisfied, so 
$\hat{t_1}\lstinline!*!$ cannot consume all matches of the expression.
We encode this constraint with the implication that $\hat{t_1}\lstinline!*!$
must match the empty string whenever $t_1\lstinline!|!\epsilon$ does.

Lookahead constrains the word to be a member of the languages of both the assertion
expression and $t_2$.
The word boundary \lstinline|\b| is effectively a single-character lookaround
for word and non-word characters.
Because the boundary can occur both ways, the model uses disjunction for the end
of $w_1$ and the start of $w_2$ being word and non-word, or non-word and word
characters, respectively. The non-word boundary \lstinline|\B| is defined as the
dual of \lstinline|\b|.

For capture groups, we bind the next capture variable $\capture{i}$ to the string
matched by $t_1$.
The $i^\text{th}$ capture group must be the 
outer capture and the remaining captures $\capture{i+1}, \ldots, \capture{i+k}$
must therefore be contained within $t_1$.
There is nothing to be done for non-capturing groups and recursion continues on
the contained subexpression.

Anchors assert the start (\lstinline!^!) and end (\lstinline[mathescape=false]|$|) 
of input; we represent the beginning and end of a word via the meta-characters $\langle$ 
and $\rangle$, respectively.
In most instances when handling these operations, $t_1$ will be $\epsilon$; 
this is because it is rare to have regex operators prior to those 
marking the start of input (or after marking the end of input, respectively).
In both these cases, we assert that the language defines the start or end of 
input---and that as a result of this, the language of $t_1$ must be an empty 
word, though the capture groups may be defined (say through $t_1$ containing 
assertions with nested captures).
We give separate rules for matching a regular expression with the multiline flag set, which
modify the behavior of anchors to accept either our meta-characters or a line break.

\subsection{Backreferences}\label{s:brefs}
\begin{table*}[tp]
\caption{Modeling backreferences.}
\label{tbl:bref-models}
\small
\begin{tabularx}{\textwidth}{lllX}
  \toprule
  \bf Type of $\backslash k$ & \bf Capturing Language & \bf Approximation & \bf Model \\

  \midrule

  Empty &
  $(w) \in \clang{\backslash k}$ &
  Exact &
  $w = \epsilon$ \\

  \midrule

  Immutable &
  $(w) \in \clang{\backslash k}$ &
  Overapproximate &
  $(\capture{k} = \varnothing \implies w = \epsilon) 
    \land (\capture{k} \neq \varnothing \implies w = \capture{k})$ \\

  \midrule

  Immutable &
  $(w) \in \clang{\backslash k\lstinline!*!}$ &
  Overapproximate &
  {$\!\begin{aligned}
    (\capture{k} = \varnothing \implies w = \epsilon)  \land
    (\capture{k} \neq \varnothing \implies  
      \exists m \geq 0 \: : \: w = \strconcat_{i=0}^{m} \capture{k})
  \end{aligned}$} \\

  \midrule

  Mutable &
  {$\!\begin{aligned}
    (w, &\capture{k}) \in 
      \clang{\lstinline[mathescape]!(?:$(t_1)\backslash k$)*!} \\
    &t_1 \text{is capture group-free}
  \end{aligned}$} &
  Overapproximate &
  {$\!\begin{aligned}
    \big(w = \epsilon \land \capture{k} & = \varnothing\big) 
    \lor \big(
      \exists m \geq 1 : 
        w = \strconcat_{i=1}^{m} (\sigma_{i,1}\strconcat\sigma_{i,2}) \\
      &\;\land
        \forall i > 1, \big( (\sigma_{i,1}, \capture{k,i}) \in \clang{t_1}
        \land \sigma_{i,2} = \capture{k,i} \big)
      \land \capture{k} = \capture{k,m}\big)  
  \end{aligned}$} \\

  \midrule

  Mutable & 
  {$\!\begin{aligned}
    (w, &\capture{k}) \in 
      \clang{\lstinline[mathescape]!(?:$(t_1)\backslash k$)*!} \\
    &t_1 \text{is capture group-free}
  \end{aligned}$} &
  Unsound &
  {$\begin{aligned}
    \big(w = \epsilon \land \capture{k} = & \varnothing\big) 
    \lor \big(
      \exists m \geq 1 : 
        w = \strconcat_{i=1}^{m} (\sigma_{i,1} \strconcat \sigma_{i,2}) \\
      &\;\land (\sigma_{i,1}, \capture{k}) \in \clang{t_1}
        \land \forall i \geq 1, (\sigma_{i,1} = \sigma_{1,1} 
          \land \sigma_{i,2} = \sigma_{1,1})
    \big)
  \end{aligned}$} \\

  \bottomrule
\end{tabularx}
\end{table*}
Table \ref{tbl:bref-models} lists our models for different cases of
backreferences in the AST of regex $R$; $\backslash k$ is a
backreference to the $k^{\text{th}}$ capture group of $R$.
Intuitively, each instance of a backreference is a variable that refers to a
capture group and has a type that depends on the structure of R.

We call a backreference \textit{immutable} if it can only evaluate to a single
value when matching; it is \textit{mutable} if it can take on multiple values,
which is a rare but particularly tricky case.
For example, consider
$\lstinline[mathescape]!/((a$|$b)$\backslash$2)+$\backslash$1$\backslash$2!/$.
Here, the backreference $\lstinline[mathescape]!$\backslash$1!$ and the second
instance of $\lstinline[mathescape]!$\backslash$2!$ are immutable.
However, the first instance of $\lstinline[mathescape]!$\backslash$2!$ is
mutable: each repetition of the outer capture group under the Kleene plus can
change the value of the second (inner) capture group, in turn changing
the value of the backreference inside this quantification.
For example, the string 
\lstinline!"aabbbbb"! satisfies this regex, but 
\lstinline!"aababaa"! does not.
To fully characterize these distinctions, we introduce the following 
definition:
\begin{defn}[Backreference Type]\label{def:types}
Let $t$ be the $k^\text{th}$ capture group of a regex $R$. Then
\begin{enumerate}
  \item $\backslash k$ is \textit{empty} if either $k$ is 
        greater than the number of capture groups in $R$, or
        $\backslash k$ is encountered before $t$ in a post-order traversal of 
        the AST of $R$;
  \item $\backslash k$ is \textit{mutable} if $\backslash k$ is not empty, and 
        both $t$ and $\backslash k$ are subterms of some quantified term $Q$ in $R$;
  \item otherwise, $\backslash k$ is \textit{immutable}.
\end{enumerate}
\end{defn}%

When a backreference is empty, it is defined as $\epsilon$, because it refers to
a capture group that either is a superterm, e.g.,
$\lstinline[mathescape]!/(a$\backslash$1)*/!$, or appears later in the
term, e.g., $\lstinline[mathescape]!/$\backslash$1(a)/!$.
 
There are two cases for immutable backreferences.
In the first case, the backreference is not quantified. %
In our model for $R$, $\capture{k}$ has already been modeled with an equality 
constraint, so we can bind the backreference to it.
In the second case, the backreference occurs within a quantification;
here, the matched word is a finite concatenation of identical copies of the
referenced capture group.
Both models also incorporate the corner case where the capture group is
$\varnothing$ due to alternation or an empty Kleene star.
Following the ES6 standard, the backreference evaluates to $\epsilon$ in
this case.

Mutable backreferences appear in the form 
$(...t_1...\backslash k...)\lstinline!*!$ where $t_1$ is the $k^\text{th}$ 
capture group; ES6 does not support forward referencing of backreferences, so 
in $(...\backslash k...t_1...)\lstinline!*!$, $\backslash k$ is empty.
For illustration purposes, the fourth entry of Table~\ref{tbl:bref-models}
describes the simplest case for mutable backreferences, other patterns are
straightforward generalizations.
In this case, we assume $t_1$ is the $k^\text{th}$ capture group but is 
otherwise capture group-free.
We can treat the entirety of this term at once: as such, any word in the 
language is either $\epsilon$, or for some number of iterations, we have the 
concatenation of a word in the language of $t_1$ followed by a copy of it.
We introduce new variables $\capture{k,i}$ referring to the values of the
capture group in each iteration, which encodes the repeated matching on the
string until settling on the final value for $\capture{k}$. %
In this instance, we need not deal with the possibility that any $\capture{k,i}$
is $\varnothing$, since the quantification ends as soon as $t_1$ does not match.

Unfortunately, constraints generated from this model are hard to solve and
not feasible for current SMT solvers, because they require ``guessing'' a
partition of the matched string variable into individual and varying components.
To make solving such queries practical, we
introduce an alternative to the previous rule where we
treat quantified backreferences as immutable. The resulting model is shown in
the last row of \autoref{tbl:bref-models}. 
E.g., returning to
$\lstinline[mathescape]!/((a$|$b)$\backslash$2)+$\backslash$1$\backslash$2!/$, 
we accept
$(\lstinline!"aaaaaaaaa"!, \lstinline!"aaaaaaaaa"!, \lstinline!"aaaa"!, 
\lstinline!"a"!)$,
but not
(\lstinline!"aabbaabbb"!, \lstinline!"aabbaabbb"!, \lstinline!"aabb"!, 
\lstinline!"b"!).
We discuss the soundness implications in \autoref{sec:enc_sound}. Quantified
backreferences are rare (see \autoref{sec:survey}), so the effect is limited in
practice.

\subsection{Modeling Non-Membership}\label{s:complement}

The model described so far overapproximates membership of a capturing language.
We define an analogous model for non-membership of the form
$\forall \capture{0}, \ldots, \capture{n}: (w, \capture{0}, \ldots, \capture{n})
\not\in \clang{R}$. Intuitively, non-membership models assert that for all capture group
assignments there exists some partition of the word such that one of the individual
constraints is violated.
Most models are simply negated. In concatenation and quantification,
only language and emptiness constraints are negated, so the models take the
form
\begin{align*}
w = w_1 & \strconcat w_2 \\ 
\land \big( & \ldots \not\in \clang{\ldots} \lor \ldots \not\in \clang{\ldots} \\
& \lor (w_2 = \epsilon \land \neg (w_1 = \epsilon \ldots))\big).
\end{align*}
In the same manner, the model for capture groups is
\[(w, \capture{i+1},..., \capture{i+k}) \not\in \clang{t_1} \land \capture{i} =
w.\]
Returning to the example of \autoref{sec:overview_model}, the negated model for
$\forall \capture{0}, \capture{1}: (w, \capture{0}, \capture{1}) \not\in \clang{\lstinline!(?:a|(b))$\backslash$1$$!}$ becomes
\begin{align*}
  &\forall \capture{0}, \capture{1} :
  w = w_1 \strconcat w'_1 \strconcat w'_2 \strconcat w_3 \\
&\land \capture{0} = w'_1 \strconcat w'_2 \\
&\land \big( \lnot \big( (w'_1 \in \lang{a} \land \capture{1} = \varnothing) \lor (w'_1 \in \lang{b} \land \capture{1} = w'_1) \big) \\
&\quad \lor \lnot (\capture{1} = \varnothing \implies w'_2 = \epsilon) \lor \lnot (\capture{1} \neq \varnothing \implies w'_2 = \capture{1}) \\
&\quad \lor w_1 \not\in \mathcal{L}(\lstinline!(:?.|\\n)*?!) \lor w_3 \not\in \mathcal{L}(\lstinline!(:?.|\\n)*?!) \big).
\end{align*}

\section{Matching Precedence Refinement}
\label{sec:refinement}
We now explain the issue of matching precedence
(\autoref{sec:matching_precedence}) and introduce a counterexample-guided
abstraction refinement scheme~(\autoref{sec:refinement_algorithm}) to address
it. We discuss termination (\autoref{sec:refinement_termination})
and the overall soundness of our approach (\autoref{sec:enc_sound}).
\subsection{Matching Precedence}
\label{sec:matching_precedence}

The model in Tables~\ref{tbl:models} and~\ref{tbl:bref-models} does not account
for matching precedence (see \autoref{sec:overview_refinement}).
A standards-compliant ES6 regex matcher will derive a unique set of
capture group assignments when matching a string $w$, because matching
precedence dictates that greedy (non-greedy) expressions match as many (as few)
characters as possible before moving on to the next~\cite{ecma-spec}. These
requirements are not part of our model, as encoding them directly into SMT would
require nesting of quantifiers for each operator, making them impractical for
automated solving.

\subsection{CEGAR for ES6 Regular Expression Models}
\label{sec:refinement_algorithm}
\newcommand{\Problem}{\ensuremath{P}}
\newcommand{\RegexList}{\ensuremath{E}}
\newcommand{\ModelName}{\ensuremath{\mathit{M}}}
\newcommand{\ModelItem}[1]{\ensuremath{\mathit{M}[#1]}}
\newcommand{\RealCapture}[1]{\ensuremath{C^{\smash{\natural}}_{#1}}}
\newcommand{\SingleRegex}[1]{\ensuremath{R_{#1}}}
\newcommand{\Refinement}{\ensuremath{R}}
\begin{algorithm}[t]
    \SetKwInOut{Input}{Input}
    \SetKwInOut{Output}{Output}
    \SetKwRepeat{Do}{do}{while}%
    \Input{
      Constraint problem \Problem{} including models for
      $m$ constraints $(w_j, \capture{0,j}, \ldots, \capture{n_j, j}) \genin_j \clang{\SingleRegex{j}}$.
    }
    \Output{
      \lstinline|null| if \Problem{} is unsatisfiable, or a satisfying assignment for \Problem{} otherwise
    }
    \SetInd{0.6em}{0.6em}
    \ModelName{} $:=$ \lstinline|null|\;
    \textit{Failed} := \lstinline|false|\;
    \Do{\textit{Failed}}{
      \ModelName{} := \lstinline!Solve!(\Problem{})\label{l:solve}\;
      \If{\ModelName{} $=$ \lstinline|null|}{
        \Return{\lstinline|null|}\label{l:unsat}\;
      }
      \textit{Failed} := \lstinline|false|\;
      \For{$j$ := $0$ \KwTo $m - 1$ \label{l:check-start}}{
        $(\RealCapture{0,j}, \ldots, \RealCapture{n_j,j})$ := \lstinline!ConcreteMatch!$(\ModelItem{w_j}, \SingleRegex{j})$ \label{l:matching}\;
        \If{$(\RealCapture{0,j}, \ldots, \RealCapture{n_j,j})$\label{l:case-start}}{
          \If{$\genin_j = \;\in$}{
            \For{$i$ := $0$ \KwTo $n_j$}{
              \If{\RealCapture{i,j} $\neq$ \ModelItem{\capture{i,j}} \label{l:yy-check}}{
                \textit{Failed} := \lstinline!true!\;
                $\Problem{} := \Problem{}  \land (w_j = \ModelItem{w_j} \!\implies\! \bigwedge_{0 \leq i \leq n_j} \capture{i,j} = \RealCapture{i,j})$ \label{l:yy-e}\;
              }
            }
          }
          \Else(\tcp*[f]{Non-membership query}){
            \textit{Failed} := \lstinline!true!\label{l:yn-s}\;
            $\Problem{} := \Problem{} \land (w_j \neq \ModelItem{w_j})$\label{l:yn-e}\;
          }
        }
        \Else(\tcp*[f]{No concrete match}){
          \If{$\genin_j = \;\in$\label{l:ny-s}}{
            \textit{Failed} := \lstinline!true!\;
            $\Problem{} := \Problem{} \land (w_j \neq \ModelItem{w_j})$\label{l:check-end}\;
          }
        }
      }
    }
    \Return{\ModelName}\label{l:done}\;
    \caption{Counterexample-guided abstraction refinement scheme for matching precedence.}
    \label{alg:refinement}
\end{algorithm}

We eliminate infeasible elements of the capturing language admitted by our 
model through counter example-guided abstraction refinement (CEGAR).

\autoref{alg:refinement} is a CEGAR-based satisfiability checker for 
constraints modeled from ES6 regexes, which relies on an external SMT solver 
with classical regular expression and string support and an ES6-compliant 
regex matcher.
The algorithm takes an SMT problem \Problem{} (derived from the DSE path
condition) as a conjunction of constraints, some of which model the
$m \geq 0$ original capturing language membership constraints.
We number the original capturing language constraints $0 \leq j < m$ so that we
can refer to them as
$(w_j, \capture{0,j}, $ $\ldots, \capture{n_j, j}) \genin_j \clang{\SingleRegex{j}}$, where
$\genin \in \{\in, \notin\}$.
The algorithm returns \lstinline!null! if \Problem{} is
unsatisfiable, or a satisfying assignment with correct matching precedence.

In a loop, we first pass the problem \Problem{} to an external SMT solver.
The solver returns a satisfying assignment \ModelName{} or
\lstinline|null| if the problem is unsatisfiable, in which case we are done
(lines \ref{l:solve}--\ref{l:unsat}).
If \ModelName{} is not null, the algorithm uses a concrete regular expression matcher (e.g.,
Node.js's built-in matcher) to populate concrete capture variables
$\capture{i,j}^\natural$ corresponding to the words $w_j$ in \ModelName{}.

Lines~\ref{l:check-start}--\ref{l:check-end} describe how 
the assignments of capture groups are checked for each regular expression 
\SingleRegex{j} in the original problem \Problem{}.
We first check whether the concrete matcher returned a list of valid capture
group assignments, i.e., whether the word $\ModelItem{w_j}$ from the satisfying
assignment matches concretely.
If it did, then $w_j$ is a member of the language generated by \SingleRegex{j}.
If $\genin_j = \;\in$, i.e., the membership constraint was positive, then we must
check if the capture group assignments are consistent with those from
\ModelName{} (line~\ref{l:yy-check}).
If they are, we move on to the next regex, otherwise we refine the constraint
problem by fixing capture group assignments to their concrete values for the
matched word (line~\ref{l:yy-e}).
Dually, if a modeled non-membership constraint was satisfiable but the word from
the current satisfying assignment $\ModelItem{w_j}$ did match concretely, we
refine the problem by asserting that $w$ must not equal that word (line
\ref{l:yn-e}).
We do the same if $\ModelItem{w_j}$ did not match concretely but came from a satisfied
positive membership constraint (line~\ref{l:check-end}).

If no refinement was necessary we have confirmed the overall assignment
satisfies \Problem{} and return \ModelName{}
(line~\ref{l:done}).
Otherwise, the loop continues with solving the refined problem. 

\subsection{Termination}
\label{sec:refinement_termination}
Unsurprisingly, CEGAR may require arbitrarily many refinements on pathological formulas and never terminate. This is unavoidable due to undecidability~\cite{bjornerStringManipulation}.
In practice, we therefore impose a limit on the number of refinements, leading
to \textit{unknown} as a possible third result. SMT solvers already may timeout
or report \textit{unknown} for complex string formulas, so this does not lead to
additional problems in practice.

\subsection{Soundness}\label{sec:enc_sound}
When constructing the rules in Tables~\ref{tbl:models}
and~\ref{tbl:bref-models}, we followed the semantics of regular expressions as
laid out in the ES6 standards document~\cite{ecma-spec}. The ES6 standard is
written in a semi-formal fashion, so we are confident that our translation into
logic is accurate, but cannot have formal proof. Existing attempts to encode
ECMAScript semantics into logic such as 
JSIL~\cite{Bodin:2014:TMJ:2535838.2535876} or KJS~\cite{Park:2015:KCF:2737924.2737991} do not 
include regexes.

With the exception of the optimized rule for mutable backreferences, our models
are overapproximate, because they ignore matching precedence.
When the CEGAR loop terminates, any spurious solutions from overapproximation
are eliminated.
As a result, we have an \textit{exact} procedure to decide (non)-membership for
capturing languages of ES6 regexes without quantified
backreferences.

In the presence of quantified backreferences, the model after CEGAR termination
becomes \textit{underapproximate}. Since DSE itself is an underapproximate
program analysis (due to concretization, solver timeouts, and partial
exploration), our model and refinement strategy are \textit{sound for DSE}.

\section{Implementation}
\label{s:impl}

We now describe an implementation of our approach in the DSE engine
ExpoSE\footnote{ExpoSE is available at \url{https://github.com/ExpoSEJS/ExpoSE}.}~\cite{Loring:2017}.
We explain how to model the regex API with capturing language
membership~(\autoref{sec:model}) and give a brief overview of
ExpoSE~(\autoref{s:expose}).

\subsection{Modeling the Regex API}\label{sec:model}

The ES6 standard specifies several methods that evaluate 
regexes~\cite{ecma-spec}. We follow its specified pseudocode for
\lstinline+RegExp.exec(s)+ to implement matching and capture group assignment 
in terms of capturing language membership in \autoref{alg:exec}.
Notably, our algorithm implements support for all flags and operators specified
for ES6.

\lstinline+RegExp.test(s)+ is precisely equivalent to the expression
\lstinline+RegExp.exec(s)+ \lstinline+!==+ \lstinline+undefined+. In the same manner, one can construct
models for other regex functions defined for ES6.
Our implementation includes partial models for the remaining functions that
allow effective test generation in practice but are not semantically complete.

\begin{algorithm}[t]
$\mathit{input}'$ := `$\langle$' + \lstinline+input+ + `$\rangle$'\;

\If{sticky {\rm\bf or} global} {
		\textit{offset} := \lstinline+lastIndex+ > 0 ? \lstinline+lastIndex+ + 1 : 0\;
    $\mathit{input}'$ := $\mathit{input}'$.\lstinline+substring+(\textit{offset})\;
}

$\mathit{source}'$ := `\lstinline!(:?.|\n)*?(!' + \lstinline+source+ + `\lstinline!)(:?.|\n)*?!';

\If{caseIgnore} {
    $source'$ := rewriteForIgnoreCase($source'$)\;
}

\If{$(\mathit{input}',\capture{0}, ..., \capture{n}) \in \clang{\mathit{source}'}$} {
  Remove $\langle$ and $\rangle$ from $(\mathit{input}', \capture{0}, ..., \capture{n})$\;
  \lstinline+lastIndex+ := \lstinline+lastIndex+ + $\capture{0}$.startIndex + $\capture{0}$.\lstinline+length+\;
  \textit{result} := [$\capture{0}, ..., \capture{n}$]\;
  \textit{result}\lstinline+.input+ := \lstinline+input+\;
  \textit{result}\lstinline+.index+ := $\capture{0}$.startIndex\;
  \Return result;
} \Else {
    \lstinline+lastIndex+ := 0\;
    \Return \lstinline+undefined+;
}

\caption{\bf{RegExp.exec(input)}}
\label{alg:exec}
\end{algorithm}

\autoref{alg:exec} first processes flags to begin from the end of the
previous match for sticky or global flags, and it rewrites the regex to
accept lower and upper case variants of characters for the ignore case flag.

We introduce the $\langle$ and $\rangle$ meta-characters to \textit{input} which act as
markers for the start and end of a string during matching.
Next, if the sticky or global flags are set we slice \textit{input} at \lstinline+lastIndex+ so
that the new match begins from the end of the previous.
Due to the introduction of our meta-characters \lstinline+lastIndex+ needs to be
offset by 1 if it is greater than zero.
We then rewrite the regex source to allow for characters to precede and 
succeed the match.
Note that we use \mbox{\lstinline!(?:.|\\n)*?!} rather than \lstinline!.*?! because
the wildcard \lstinline!.! consumes all characters except line breaks in 
ECMAScript regexes.
To avoid adding these characters to the final match we place the original 
regex source inside a capture group. This forms $\capture{0}$, which is 
defined to be the whole matched string~\cite{ecma-spec}. 
Once preprocessing is complete we test whether the input string and fresh 
string for each capture group are within the capturing language for the 
expression.
If they are then a results object is created which returns the correctly 
mapped capture groups, the input string, and the start of the match in the 
string with the meta-characters removed. Otherwise \lstinline+lastIndex+ is
reset and \lstinline|undefined| is returned.

\subsection{ExpoSE}
\label{s:expose}

ExpoSE is a DSE engine which uses the Jalangi2~\cite{Gong:2015:DDC:2771783.2771809} framework to 
instrument a piece of JavaScript software in order to create a program trace.
As the program terminates, ExpoSE calls the SMT solver Z3~\cite{Z3Solver} to 
identify all feasible alternate test-cases from the trace.
These new test cases are then queued and the next test case is selected for
execution, in the manner of generational search~\cite{AWBFT:Godefroid}.
The ExpoSE framework allows for the parallel execution of individual test 
cases, aggregating coverage and alternative path information as each test case 
terminates.
This parallelization is achieved by executing each test case as a unique 
process allocated to a dedicated single core; as such the analysis is highly 
scalable.

Our strategy for test case selection is similar to the CUPA strategy proposed by~\citet{CUPAStrat}.
We use program fork points to prioritize unexplored code: each expression is given a unique identifier and scheduled test cases are sorted into buckets based upon which expression was being executed when they were created.
We select the next test case %
by choosing a random test case
from the bucket that has been accessed least during the analysis; this 
prioritizes test cases triggered by less common expressions.

\section{Evaluation}
\label{sec:eval}
We now empirically answer the following research questions:
\begin{enumerate}[label=(\textbf{RQ\arabic*}),align=parleft,leftmargin=*]

\item \label{item:rq1} Are non-classical regexes an important
  problem in JavaScript?

\item \label{item:rq2} Does accurate modeling of ES6 regexes make
  DSE-based test generation more effective?

\item \label{item:rq3} Does the performance of the model and the refinement
  strategy enable practical analysis?

\end{enumerate}
We answer the first question with a survey of regex usage in the
wild (\autoref{sec:survey}).
We address \hyperref[item:rq2]{RQ2} by comparing our approach against an
existing partial implementation of regex support in
ExpoSE~\cite{Loring:2017} on a set of widely used libraries
(\autoref{sec:prev_comparison}).
We then measure the contribution of each aspect of our approach on over 1,000
JavaScript packages~(\autoref{sec:efficacy}).
We answer \hyperref[item:rq3]{RQ3} by analyzing solver and refinement 
statistics per query (\autoref{sec:rwre}).

\subsection{Surveying Regex Usage}
\label{sec:survey}
We focus on code written for Node.js, a popular framework for standalone
JavaScript. Node.js is used for both server and desktop applications, including
popular tools \textit{Slack} and \textit{Skype}.
We analyzed 415,487 packages from the NPM repository, the primary software 
repository for open source Node.js code.
Nearly 35\% of NPM packages contain a regex,
20\% contain a capture group and 4\% contain a backreference.

\paragraph{Methodology}
We developed a lightweight static analysis that parses all source files in a 
package and identifies regex literals and function calls.
We do not detect expressions of the form \mbox{\lstinline!new RegExp(...)!,} as
they would generally require a more expensive static analysis. 
Our numbers therefore provide a lower bound for regex usage.

\paragraph{Results}

\begin{table}[t]

\caption{Regex usage by NPM package.}
\label{tab:mod_occurence}

\begin{tabularx}{\columnwidth}{X|rr}
\toprule
\bf Feature & \bf Count & \% \\
\midrule
Packages on NPM & 415,487 & 100.0\% \\
$\;\;\ldots$ with source files & 381,730 & 91.9\% \\
$\;\;\ldots$ with regular expressions & 145,100 & 34.9\% \\
$\;\;\ldots$ with capture groups & 84,972 & 20.5\% \\
$\;\;\ldots$ with backreferences & 15,968 & 3.8\% \\
$\;\;\ldots$ with quantified backreferences & 503 & 0.1\% \\
\bottomrule
\end{tabularx}
\end{table}

\begin{table}[t]

\caption{Feature usage by unique regex.}
\label{tab:regex_occurence}

\begin{tabularx}{\columnwidth}{X|rr|rr}
\toprule
\bf Feature & \bf Total & \% & \bf Unique \hspace*{-4mm} & \% \\
\midrule
Total Regex & 9,552,546 & 100\% & 305,691 & 100\% \\
Capture Groups & 2,360,178 & 24.71\% & 119,051 & 38.94\% \\
Global Flag & 2,620,755 & 27.44\% & 90,356 & 29.56\% \\
Character Class & 2,671,565 & 27.97\% & 71,040 & 23.24\% \\ 
Kleene+ & 1,541,336 & 16.14\% & 67,508 & 22.08\% \\  
Kleene* & 1,713,713 & 17.94\% & 66,526 & 21.76\% \\ 
Ignore Case Flag & 1,364,526 & 14.28\% & 58,831 & 19.25\% \\  
Ranges & 1,273,726 & 13.33\% & 52,155 & 17.06\% \\ 
Non-capturing & 1,236,533 & 12.94\% & 25,946 & 8.49\% \\ 
Repetition & 360,578 & 3.7\% & 17,068 & 5.58\% \\ 
Kleene* (Lazy) & 230,060 & 2.41\% & 13,250 & 4.33\% \\ 
Multiline Flag & 137,366 & 1.44\% & 10,604 & 3.47\% \\ 
Word Boundary & 336,821 & 3.53\% & 9,677 & 3.17\% \\ 
Kleene+ (Lazy) & 148,604 & 1.56\% & 6,072 & 1.99\% \\ 
Lookaheads & 176,786 & 1.85\% & 3,123 & 1.02\% \\ 
Backreferences & 64,408 & 0.67\% & 2,437 & 0.80\% \\ 
Repetition (Lazy) & 2,412 & 0.03\% & 221 & 0.07\% \\ 
Quantified BRefs & 1,346 & 0.01\% & 109 & 0.04\% \\ 
Sticky Flag & 98 & <0.01\% & 60 & 0.02\% \\ 
Unicode Flag & 73 & <0.01\% & 48 & 0.02\% \\ 
\bottomrule
\end{tabularx}
\end{table}

We found regex usage in JavaScript to be widespread, with 
145,100 packages containing at least one regex out of a total 
415,487 scanned packages.
\autoref{tab:mod_occurence} lists the number of NPM packages 
containing regexes, capture groups, backreferences, and %
backreferences appearing within quantification.
Note that a significant number of packages make use of capture groups and 
backreferences, confirming the importance of supporting them.

\autoref{tab:regex_occurence} reports statistics for all 9M regexes
collected, giving for each feature the fraction of expressions including it.
Many regexes in NPM packages are not unique;
this appears to be due to repeated inclusion of the same literal (instead of
introduction of a constant), the use of online solutions to common problems,
and the inclusion of dependencies (foregoing proper dependency management). 
To adjust for this, we provide data for both all expressions encountered and 
for just unique expressions.
In both cases, there are significant numbers of capture groups, backreferences,
and other non-classical features.
As the occurrence rate of quantified backreferences is low, we do not 
differentiate between mutable and immutable backreferences.

\paragraph{Conclusions}
Our findings confirm that regexes are widely used and often contain
complex features.
Of particular importance is a faithful treatment of capture groups, which 
appear in $20.45\%$ of the packages examined.
On the flip side, since quantified backreferences make up just
$0.01\%$ of regexes, the optimization introduced in \autoref{s:brefs} will
rarely lead to additional underapproximation during DSE.

\subsection{Improvement Over State of the Art}
\label{sec:prev_comparison}

We compare our approach against the original ExpoSE~\cite{Loring:2017}, which
is, to our knowledge, the only available and functional implementation of regex
support in JavaScript.

\paragraph{Methodology}

We evaluated %
statement coverage achieved by both versions of ExpoSE on a set
of libraries, which we chose for their popularity (with up to 20M weekly
downloads) and use of regex. This includes the three libraries
minimist, semver, and validator, which the first version of ExpoSE was
evaluated on \cite{Loring:2017}.
To fairly compare original ExpoSE against our extension, we use the original
automated library harness for both. Therefore we do not
take advantage of other improvements for test generation, such as symbolic array
support, which we have added in the course of our work.
We re-executed each package six times for one hour each on both versions, using
32-core machines with 256GB of RAM, and averaged the results.
We limited the refinement scheme to 20 iterations, which we identified as
effective in preliminary testing (see \autoref{sec:rwre}).

\paragraph{Results}
\begin{table}[t]
  \caption{Statement coverage with our approach (\textbf{New}) vs. \cite{Loring:2017}
    (\textbf{Old}) and the relative increase (\textbf{+}) on popular NPM
    packages (\textbf{Weekly} downloads). \textbf{LOC} are lines loaded and
    \textbf{RegEx} are regular expression functions symbolically
    executed.}
  \label{tbl:prev_work_comparison_tbl}
  \small
  \begin{tabular}{l@{\hspace*{-1mm}}S[table-format=5.0]@{\hspace*{3mm}}S[table-format=5.0]@{\hspace*{2mm}}S[table-format=5.0]@{\hspace*{1mm}}S[table-format=2.1]@{\hspace*{1mm}}S[table-format=2.1]@{\hspace*{0mm}}S[table-format=4.1]}
    \toprule
    {\bf Library} & {\bf Weekly} & {\bf LOC} & {\bf RegEx} & {\bf Old{\footnotesize(\%)}} & {\bf New{\footnotesize(\%)}} & {\bf \ \ \ +{\footnotesize(\%)}} \\
    \midrule
    babel-eslint    & 2500k & 23047 &   902 & 21.0 & 26.8 & 27.6 \\
    fast-xml-parser & 20k   &   706 &   562 &  3.1 & 44.6 & 1338.7 \\
    js-yaml         & 8000k &  6768 &    78 &  4.4 & 23.7 & 438.6 \\
    minimist        &20000k &   229 & 72530 & 65.9 & 66.4 & 0.8 \\
    moment          & 4500k &  2572 &    21 &  0.0 & 52.6 & {$\;\;\;\;\;\infty$} \\
    query-string    & 3000k &   303 &    50 &  0.0 & 42.6 & {$\;\;\;\;\;\infty$} \\
    semver          & 1800k &   757 &   616 & 51.7 & 46.2 & -10.6 \\ 
    url-parse       & 1400k &   322 &   448 & 60.9 & 71.8 & 17.9 \\
    validator       & 1400k &  2155 &    94 & 67.5 & 72.2 & 7.0 \\
    xml             & 500k  &   276 &  1022 & 60.2 & 77.5 & 28.7 \\
    yn              & 700k  &   157 &   260 &  0.0 & 54.0 & {$\;\;\;\;\;\infty$} \\
    \bottomrule
  \end{tabular}
\end{table}
\autoref{tbl:prev_work_comparison_tbl} contains the results of our
comparison. 
To provide an indication of program size, we use the number of lines of code
loaded at runtime (JavaScript's dynamic method of loading dependencies makes it
hard to determine a meaningful LOC count statically).

The results demonstrate that ExpoSE extended with our model and refinement strategy 
can improve coverage more than tenfold on our sample of widely-used libraries.
In the cases of moment, query-string, and yn, the lack of ES6 support in the
original ExpoSE prohibited meaningful analysis, leading to 0\% coverage.
In the case of semver, we see a decrease in coverage if stopped after one
hour. This is due to the modeling of regex increasing
solving time (see also \autoref{sec:rwre}). The coverage deficit disappears 
when executing both versions of ExpoSE with a timeout of two hours.

\paragraph{Conclusions}
We find that our modifications to ExpoSE make test generation more effective in
widely used libraries using regex.
This suggests that the new method of solving regex queries
presented in this paper has a substantial impact on practical problems in DSE.
We also see that other improvements to ExpoSE, such as ES6 support, have
affected coverage. Therefore, we continue with an evaluation of the individual
aspects of our model.

\subsection{Breakdown of Contributions}
\label{sec:efficacy}
We now drill down into how the individual improvements in regex support are
contributing to increases in coverage.

\paragraph{Methodology}
From the packages with regexes from our survey 
\autoref{sec:survey}, we developed a test suite of 1,131 NPM libraries for 
which ExpoSE is able to automatically generate a meaningful test harness.
In each of the libraries selected, ExpoSE executed at least one regex 
operation on a symbolic string, which ensures that the library
contains some behavior relevant to the scope of this paper.
The test suite constructed in this manner contains numerous libraries that are
dependencies of packages widely used in industry, including Express
and Lodash.\footnote{Raw data for the experiments, including all
  package names, is available at \url{https://github.com/ExpoSEJS/PLDI19-Raw-Data}.}

Automatic test generation typically requires a bespoke test harness or set of
parameterized unit tests~\cite{TillmannS05a} to achieve high coverage in code
that does not have a simple command line interface, including libraries.
ExpoSE's harness explores libraries fully automatically by executing all exported
methods with symbolic arguments for the supported types \lstinline|string|,
\lstinline|boolean|, \lstinline|number|, \lstinline|null| and
\lstinline|undefined|.
Returned objects or functions are also subsequently explored in the same manner.

We executed each package for one hour, which typically allowed to reach a
(potentially initial) coverage plateau, at which additional test cases do not
increase coverage further.
We break down our regex support into four levels and measure the
contribution and cost of each one to line coverage and test execution rate
(\autoref{tbl:cov_delta}).
As baseline, we first execute all regex methods concretely,
concretizing the arguments and results. In the second configuration, we add the
model for ES6 regex and their methods, including support for word
boundaries and lookaheads, but remove capture groups and concretize any 
accesses to them, including backreferences. Third, we also enable full support 
for capture groups and backreferences. Fourth, we finally also add the 
refinement scheme to address overapproximation.

\paragraph{Results}
\begin{table}[tp]
  \caption{Breakdown of how different components contribute to
    testing $1,131$ NPM packages, showing number (\textbf{\#}) and fraction
    (\textbf{\%}) of packages with coverage improvements, the geometric mean of the relative coverage
    increase from the feature (\textbf{Cov}), and test execution rate.}
\label{tbl:cov_delta}
\small
\begin{tabularx}{\columnwidth}{X|rr|r|rr}
  \toprule
  &  \multicolumn{2}{c|}{\bf Improved} & \multicolumn{1}{c|}{\bf Cov} &  \\
  \bf Regex Support Level & \multicolumn{1}{c}{\bf \#} & \multicolumn{1}{c|}{\bf \%} & \multicolumn{1}{c|}{\bf +(\%)} & \smash{\raisebox{2.5pt}{$\frac{\textbf{Tests}}{\textbf{min}}$}} \\
  \midrule
  Concrete Regular Expressions & - & - & - & 11.46 \\
  + Modeling RegEx & 528 & 46.68\% & +6.16\% & 10.14 \\
  + Captures \& Backreferences & 194 & 17.15\% & +4.18\% & 9.42 \\
  + Refinement & 63 & 5.57\% & +4.17\% & 8.70 \\
\midrule
    All Features vs. Concrete & 617 & 54.55\% & +6.74\% & \\
  \bottomrule
\end{tabularx}
\end{table}
\autoref{tbl:cov_delta} shows, for each level of support, the number and
percentage of target packages where coverage improved; the geometric mean of 
the relative increase in coverage; and the mean test execution rate.
The final row shows the effect of enabling full support compared to the
baseline. Note that the number of packages improved is less than the sum of the
rows above, since the coverage of a package can be improved by multiple
features.

In a dataset of this size that includes many libraries that make only little 
use of regex, average coverage increases are expected to be
small. Nevertheless, we see that dedicated support improves the coverage of 
more than half of packages that symbolically executed at least one regex 
function.
As expected, the biggest improvement comes from supporting basic symbolic
execution of regular expressions, even without capture groups or regard for
matching precedence. However, we see further improvements when adding capture
groups, which shows that they indeed affect program semantics.
Refinement affects fewer packages, although it significantly contributes to coverage where it is required.
This is because a lucky solver may generate correct inputs on
the first attempt, even in ambiguous settings.

On some libraries in the dataset, the approach is highly effective.
For example, in the manifest parser \textit{n4mf-parser}, full support improves
coverage by $29\%$ over concrete; in the format conversion library
\textit{sbxml2json}, by $14\%$; and in the browser detection library
\textit{mario}, by $16\%$.
In each of these packages the refinement scheme contributed to the improvement
in coverage.
In general, the largest increases are seen in packages that include
regular expression-based parsers.

Each additional feature causes a small decrease in average test execution rate.
Although a small fraction ($\sim$1\%) of queries can take longer than 300s to
solve, concurrent test execution prevents DSE from stalling on a single query.

\paragraph{Conclusions}
Full support for ES6 regex improves performance of DSE of
JavaScript in practice at a cost of a $16\%$ increase in execution
time~(\hyperref[item:rq2]{RQ2}).
An increase in coverage at lower execution rate in a fixed time window suggests
that full regular expression support increases the quality of individual test
cases.

\subsection{Effectiveness on Real-World Queries}
\label{sec:rwre}
We now investigate the performance of the model and refinement scheme to answer
\hyperref[item:rq3]{RQ3}.
Finally, we also discuss the refinement limit and how it affects analysis.

\paragraph{Methodology}
We collected data on queries during the NPM experiments (\autoref{sec:efficacy})
to provide details on SMT query success rates and execution times, as well as
on the usage of the refinement scheme.

\paragraph{Results}
\begin{table}[t]
\caption{Solver times per package and query.}
\label{tab:query_stats}
\small%
\begin{tabularx}{\columnwidth}{@{\hspace*{1pt}}l@{\hspace*{1pt}}rrr}
  \toprule
  & \multicolumn{3}{c}{\bf Constraint Solver Time} \\
\bf Packages/Queries & \bf Minimum & \bf Maximum & \bf Mean \\
\midrule
All packages & 0.04s & 12h 15m & 2h 34m \\
With capture groups & 0.20s & 12h 15m & 2h 40m \\
With refinement  & 0.46s & 12h 15m & 2h 48m \\
\hspace*{-0.75mm} Where refinement limit is hit & 3.49s & 11h 07m & 3h 17m \\
\midrule
All queries & 0.001s & 22m 26s & 0.15s \\
With capture groups & 0.001s & 22m 26s & 5.53s \\
With refinement & 0.005s & 18m 51s & 22.69s \\
Where refinement limit is hit & 0.120s & 18m 51s & 58.85s \\
\bottomrule
\end{tabularx}
\end{table}
We found that 753 (66\%) of the 1,131 packages tested executed at least one query containing a capture group or backreference.
Of these packages, 653 (58\% overall) contained at least one query to the SMT 
solver requiring refinement, and 134 (12\%) contained a query that 
reached the refinement limit.

In total, our experiments executed 58,390,184 SMT queries to generate test
cases.
As expected, the majority do not involve regexes, but they form a 
significant part:
4,489,581 (7.6\%) queries modeled a regex, 645,295 (1.1\%) 
modeled a capture group or backreference, 74,076 (0.1\%) required use of the  
refinement scheme and 2,079 (0.003\%) hit the refinement limit.
The refinement scheme was overwhelmingly effective: only $2.8\%$ of queries with
at least one refinement also reached the refinement limit ($0.003\%$ of all
queries where a capture group was modeled).
Of the refined SMT queries, the mean number of refinements required to produce 
a valid %
satisfying assignment was 2.9; the majority of queries required 
only a single refinement.

\autoref{tab:query_stats} details time spent processing SMT problems per-package and per-query. 
We provide the data over the four key aspects of the problem: we report the time spent in the constraint solver both per package and per query in total, as well as the time in the constraint solver for the particularly challenging parts of our strategy.
We found that the use of refinements increased the average per-query solving
time by a factor of four; however, this is dominated by SMT queries that hit the
refinement limit, which took ten times longer to run on average.
The low minimum time spent in the solver in some packages can be attributed to
packages where a regular expression was encountered early in execution but
limitations in the test harness or function models (unrelated to regular
expressions) prevented further exploration.

\paragraph{Conclusions}

We find the refinement scheme is highly effective, as it is able to solve 97.2\%
of encountered constraint problems containing regexes.
It is also necessary, as $10\%$ of queries containing a capture group had led to a spurious satisfying assignment and required refinement.

Usually, only a small number of refinements are required to produce a correct
satisfying assignment. Therefore, even refinement limits of five or fewer are
feasible and may improve performance with low impact on coverage.

\subsection{Threats to Validity}

We now look at potential issues affecting the validity of our results, in
particular soundness, package selection, and scalability.

\paragraph{Soundness}

In addition to soundness of the model (see \autoref{sec:enc_sound}), one must
consider the soundness of the implementation.  In the absence of a mechanized
specification for ES6 regex, our code cannot be proven correct, so we use an extensive
test suite for validation.
However, assuming the concrete matcher is specification-compliant,
\autoref{alg:refinement} will, if it terminates, return a
specification-compliant model of the constraint formula even if the
implementation of \autoref{sec:encoding} contains bugs. In the worst case, the
algorithm would not terminate, leading to timeouts and loss of coverage.
Bugs could therefore only have lowered the reported
coverage improvements.

\paragraph{Package Selection and Harness}

In \autoref{sec:efficacy}, we chose packages identified in our survey
(\autoref{sec:survey}) where our generic harness encountered a regular
expression within one hour of DSE. This allowed us to focus the evaluation on
regex support as opposed to evaluating the quality of the harness (and having 
to deal with unreachable code in packages).
Use of this harness may have limited package selection to simpler, 
unrepresentative libraries.
However, we found that simple APIs do not imply simple code: the final dataset
contains several complex packages, such as language parsers, and the types of
regexes encountered were in line with the survey results.
On simple code we found that ExpoSE would often reach 100\% coverage;
failure to do so was either due to the complexity of the code or the lack of
support for language features unrelated to regex and APIs that would require
additional modeling (e.g., the file system).

\paragraph{Scalability}
Scalability is a challenge for DSE in general, and is not specific to our 
model for regex.
Empirically, execution time for a single test (instrumentation, execution,
and constraint generation) grows linearly with program size, as does the average 
size of solver queries.
The impact of query length on solving time varies, but does not appear to be
exacerbated by our regex model.
In principle, our model is compatible with compositional
approaches~\cite{Godefroid:2007:CDT:1190216.1190226,
  Anand:2008:DCS:1792734.1792771} and state
merging~\cite{statemerging-pldi12,veritesting}, which can help DSE scale to
large programs.

The scalability of our approach suffices for Node.js, however: JavaScript has
smaller LOC counts than, e.g., C++, and code on NPM is very modular.
For instance, among the top 25 most depended-upon NPM libraries, the
largest is 30 KLOC (but contains no regex).
Several packages selected for our evaluation, such as babel-eslint, had 
between 20-30 KLOC and were meaningfully explored with the generic 
harness.
 
\section{Related Work}
\label{sec:relatedwork}

In prior work, we introduced ExpoSE and partial support for encoding JavaScript
regex in terms of classical regular language membership and string
constraints~\cite{Loring:2017}. This initial take on the problem was lacking
support for several problematic features such as lookaheads, word boundaries,
and anchors. Matching precedence was presented as an open problem, which we have
now addressed through our refinement scheme.

In theory, regex engines can be symbolically executed themselves
through the interpreter~\cite{CUPAStrat}. While this removes the need for
modeling, in practice the symbolic execution of the entire interpreter and
regex engine quickly becomes infeasible due to path explosion.

There have been several other approaches for symbolic execution of Java\-Script; %
most include some limited support for classical regular expressions.
\citet{li2009reggae} presented an automated test generation scheme for
programs with regular expressions by on-line generation of a matching function
for each regular expression encountered, exacerbating path explosion.
\citet{Saxena:2010:SEF:1849417.1849985} proposed the first scheme
to encode capture groups through string constraints.
\citet{jalangi} presented Jalangi, a tool based on program
instrumentation and concolic values.
\citet{pass:Li} and~\citet{symjs:Li} describe a custom browser and symbolic
execution engine for JavaScript and the browser DOM, and a string constraint
solver \textit{PASS} with support for most JavaScript string
operations. 
Although all of these approaches feature some support for ECMAScript 
regex (such as limited support for capture groups), they  
ignore matching precedence and do not support backreferences or lookaheads.

\citet{DBLP:conf/icse/ThomeSBB17} propose a heuristic approach for 
solving constraints involving unsupported string operations.
We choose to model operations unsupported by the solver and employ a CEGAR
scheme to ensure correctness.
\citet{Abdulla:2017:FCF:3062341.3062384} propose the use of a
refinement scheme to solve complex constraint problems, including support for
context-free languages.
The language of regular expressions with backreferences is not
context-free~\cite{doi:10.1142/S012905410300214X} and, as such, their scheme
does not suffice for encoding all regexes; however, their approach
could serve as richer base theory than classic regular expressions.
\citet{Scott2015} suggest backreferences can be eliminated via concatenation 
constraints, however they do not present a method for doing so.

Further innovations from the string solving community, such as work on the 
decidability of string constraints involving complex 
functions~\cite{DBLP:journals/pacmpl/ChenCHLW18, DBLP:journals/pacmpl/HolikJLRV18}
or support for recursive string 
operations~\cite{DBLP:conf/cav/TrinhCJ16, DBLP:conf/cav/TrinhCJ17},
are likely to improve the performance of our approach in future.
We incorporate our techniques at the level of the DSE engine rather than the
constraint solver, which allows our tool to leverage advances in string solving
techniques; at the same time, we can take advantage of the native regular
expression matcher and can avoid having to integrate implementation
language-specific details for regular expressions into the solver.

A previous survey of regex usage across 4,000 Python 
applications~\cite{Chapman:2016:ERE:2931037.2931073} also provides a strong 
motivation for modeling regex.
Our survey extends this work to JavaScript on a significantly larger sample
size.

\section{Conclusion}
\label{sec:conclusion}
In this paper we presented a model for the complete regex 
language of ES6, which is sound for the dynamic symbolic execution of the 
\lstinline!test! and \lstinline!exec! functions.
We model regex membership constraints in terms of string constraints and 
classical regular language membership.
We introduced a novel CEGAR scheme to address the challenge of matching 
precedence, which so far had been largely ignored in related work.
To the best of our knowledge, ours is the first comprehensive solution for
ES6.
We demonstrated that regexes---and specifically their non-regular features---are 
extensively used in JavaScript and that existing DSE-based analyses would 
therefore suffer coverage loss from concretization.
In a large scale evaluation of over 1,000 Node.js programs, our novel solution 
outperforms existing partial approaches to the problem and demonstrates the 
viability of our model for improving the analysis of string-manipulating 
JavaScript programs.

\begin{acks}
  Blake Loring was supported by the EPSRC \grantsponsor{epsrc}{Centre for
  Doctoral Training in Cyber Security at Royal Holloway, University of
  London}{https://www.epsrc.ac.uk} (\grantnum{epsrc}{EP/K035584/1}).
\end{acks}

\bibliography{bibliography}

\end{document}